\documentclass[aps,pra,twocolumn,showpacs,superscriptaddress,groupedaddress]{revtex4-1}
\usepackage{graphicx}
\usepackage{epstopdf}
\usepackage{amsmath}
\usepackage{amssymb}
\usepackage{textcomp}
\usepackage{xspace}
\usepackage{natbib}
\newcommand{\ket}[1]{\ensuremath{|#1\rangle}\xspace}

\begin{document}
%\title{Fast and Robust Microwave Beam-Splitter For Thermal Atom Interferometer On Chip}
\title{Symmetric micro-wave potentials for interferometry with thermal atoms on a chip}

%\author{}
%\email{mahdi.ammar@thalesgroup.com}
%\thanks{}
\author{M. Ammar$^{1,2}$, M. Dupont-Nivet$^{1,3}$, L. Huet$^{1}$, J.-P. Pocholle$^{1}$, P.~Rosenbusch$^{4}$, I.~Bouchoule$^{3}$, C. I. Westbrook$^{3}$, J.~Est\`{e}ve$^{2}$, J.~Reichel$^{2}$, C.~Guerlin$^{2}$ and S. Schwartz$^{1}$}
\affiliation{${}^{1}$Thales Research and Technology France, Campus Polytechnique, 1 av. Fresnel, 91767 Palaiseau, France \\
${}^{2}$Laboratoire Kastler-Brossel, ENS, CNRS, Universit\'e Pierre et Marie Curie-Paris 6, 24 rue Lhomond, 75005 Paris, France \\
${}^{3}$Laboratoire Charles Fabry de l'Institut d'Optique, Campus Polytechnique, 2 av. Fresnel, 91127 Palaiseau, France \\
${}^{4}$LNE-SYRTE, Observatoire de Paris, UPMC, CNRS, 61 av de l'Observatoire, 75014 Paris, France}

\date{\today}
%
%%%%%%%%%%%%%%%%%%%%%%%%%%%%%%%%%%%%%%%%%%%%%%%%%%%%%%%%%%%
\begin {abstract}
A trapped atom interferometer involving state-selective adiabatic potentials with two microwave frequencies on a chip is proposed. We show that this configuration provides a way to achieve a high degree of symmetry between the two arms of the interferometer, which is necessary for coherent splitting and recombination of thermal (i.e. non-condensed) atoms. The resulting interferometer holds promise to achieve high contrast and long coherence time, while avoiding the mean-field interaction issues of interferometers based on trapped Bose-Einstein condenstates.
\end {abstract}
%\keywords{}
\maketitle

% body of paper here - Use proper section command
\section{INTRODUCTION}
\label{SecIntro}
%%%%%%%%%%%%%%%%%%%%%%%%%%%%%%%%%%%%%%%%%%%%%%%%%%%%%%%%%%%%%%%%%%%%%%%%%%%%%%%%%%
%
Atom interferometers \cite{cronin2009optics} have proven very successful in precision measurements such as the determination of the fine structure constant \cite{wicht_2002,bouchendira2011new}, the determination of the Newtonian gravitational constant \cite{lamporesi_2008}, and inertial sensing of gravity \cite{peters_2001}, gravity gradients \cite{mcguirk_2002} and rotation \cite{gustavson2000rotation}. They also show great promise to perform general relativity tests \cite{dimopoulos_2008_b}, including the weak equivalence principle \cite{bonnin_2013,Tino_2013}.

In parallel, atom chips \cite{reichel1999atomic, folman2000controlling, fortagh2007magnetic,reichel2010atom} provide a robust and versatile tool to trap and manipulate ultracold atoms, and are now routinely used in a variety of setups, including free-falling experiments in a drop tower \cite{van_2010}, and compact atomic clocks \cite{Maineult2012}. In this context, they are very promising candidates for next-generation compact atomic sensors, including onboard applications \cite{zatezalo2008bose}. However, while a variety of integrated beam splitters and coherent manipulation techniques have been demonstrated \cite{shin_2005,schumm2005matter,treutlein2006microwave,jo2007long,bohi2009coherent,maussang2010,Hinds2010,berrada_2013}, none of the chip-based atom interferometers developed so far has reached metrological performance.

One of the initial problems encountered by atom-chip interferometers, namely the difficulty to maintain stable trapping and a reasonable trap-surface distance during the coherent splitting process \cite{davis_2002,shin_2005}, has been overcome by the use of dressed state adiabatic potentials \cite{zobay2001,schumm2005matter}. However, another issue remains unresolved: trapped-atom interferometers using Bose-Einstein condensates (BECs) are especially sensitive to atom-atom interactions which induce phase diffusion, limiting their coherence time  \cite{lewenstein1996quantum, javanainen1997phase, grond2010atom} and putting a serious constraint on the achievable precision in the measurement of the relative phase between the two arms of the interferometer \cite{schumm2005matter,Hinds2010}.

One possible way to reduce the interaction strength, that we investigate throughout this paper, is the use of a trapped but thermal (i.e. non-degenerate) atomic cloud whose density is sufficiently low that the effect of interactions is negligible. This choice is somewhat analogous to using incoherent light in an optical interferometer, as already pointed out in \cite{andersson2002multimode} for guided thermal atoms propagating through two combined Y-shaped beam splitters. As in a "white light interferometer", the short coherence length of a thermal cloud (typically the thermal de Broglie wavelength \cite{cohen2011advances}) requires that the interferometer be kept sufficiently symmetric (in a sense that will be defined in section \ref{SecMW}) in order to observe any interference.

With this aim in view, we propose a protocol for a symmetric atom interferometer suitable for thermal atoms, using internal state labeling and adiabatic dressed potentials based on the same principle as in \cite{bohi2009coherent}. In the work of Ref.~\cite{bohi2009coherent}, which involves BECs, only one of the two internal states is dressed, breaking the spatial symmetry of the interferometer because the microwave field renders the trapping frequencies different for the two interferometer paths. To restore the symmetry, we propose the use of \emph{two} microwave frequencies on \emph{two} separate planar waveguides, each one interacting (primarily) with one of the two internal states. Thus each interferometer path can be individually controlled and made nearly identical to the other.

This paper is organized as follows: we first discuss and quantify the role of symmetry in terms of interferometer contrast; we then describe the basic principles of the proposed protocol, and show why it is well suited for achieving a symmetric configuration; we then compare attractive and repulsive microwave fields, and show that the latter are much more favorable in this context; finally, taking into account how the atomic energy levels are affected by the presence of both static and microwave fields, we discuss the robustness of the design against external field fluctuations.

%
%%%%%%%%%%%%%%%%%%%%%%%%%%%%%%%%%%%%%%%%%%%%%%%%%%%%%%%%%%%%%%%%%%%%%%%%
\section{Role of symmetry in the interferometer contrast}
\label{SecMW}
%%%%%%%%%%%%%%%%%%%%%%%%%%%%%%%%%%%%%%%%%%%%%%%%%%%%%%%%%%%%%%%%%%%%%%%%%%%%%
%
To model a trapped atom interferometer, let us consider an ensemble of atoms, each one having two internal states labelled $\left|a \right>$ and $\left|b \right>$ (a possible practical realization will be discussed in the next section). The atoms are supposed to be initially prepared in the internal state $\left|a\right>$, at thermal equilibrium with temperature $T$ in the trapping potential $V_a$. The temperature $T$ is moreover assumed to be high enough that Boltzmann statistics apply (for bosons in a harmonic trap with a BEC transition temperature $T_c$, this means that $T$ is at least on the order of a few $T_c$, so that the gas is only weakly degenerate \cite{pethick2002bose}). The atoms are first put into a coherent superposition of $\ket{a}$ and $\ket{b}$ with equal weight by applying a $\pi/2$ pulse (described by the unitary operator $\exp{\left(-i\pi\widehat{\sigma}/4\right)}$, where $\widehat{\sigma}=\left|a\right>\left<b\right| + \left|b\right>\left<a\right|$). Then the two internal states are spatially separated and recombined using state-dependent potentials $V_{a,b}(\widehat{z},t)$ (where $\widehat{z}$ is the position operator and $t$ the time), with $V_a(\widehat{z},t_i)=V_a(\widehat{z},t_f)=V_b(\widehat{z},t_i)+C=V_b(\widehat{z},t_f)+C$, where $t_i$ and $t_f$ are respectively the initial and final time of the sequence, and $C$ is a constant energy term. Finally, another $\pi/2$ pulse is applied to convert the phase difference into a population difference, following a typical Ramsey sequence.

Between the two $\pi/2$ pulses, the evolution of the system is governed by the following Hamiltonian:
\begin{equation} \label{hamiltonian}
\widehat{H}= \widehat{p}^2/2m + V_{a}(\widehat{z},t) \left|a\right>\left<a\right| + V_b(\widehat{z},t) \left|b\right>\left<b\right| \;,
\end{equation}
where $\widehat{p}$ is the momentum operator. In this formalism, the evolution of an atom in the internal state $\left|a \right>$ will be governed by the Hamiltonian $\widehat{H} \left|a \right> \left< a \right|$, with eigenenergies $E_n^a$ and eigenvectors $\left|n\right>_a \left|a\right>$, where $\left|n\right>_a$ refer to the external state of the atom. To simplify the notations in the following, we introduce $\left|n,a\right>=\left|n\right>_a \left|a\right>$, and its counterpart $\left|n,b\right>=\left|n\right>_b \left|b\right>$ for the eigenvectors of $\widehat{H} \left|b \right>\left< b \right|$. Before the first $\pi/2$ pulse, the atomic cloud can thus be described by the density matrix $\widehat{\rho} = \sum_n p_n \left|n,a\right>\left<n,a\right|$, where $p_n=e^{-E_n^a/kT}/Z$ are the Boltzmann factors and $Z=\sum_n e^{-E_n^a/kT}$ is the partition function.

Throughout this paper, we neglect the effects of collisions and assume in particular that the atomic ensemble does not have time to re-thermalize during the interferometric sequence. Furthermore, we only consider the case in which the two internal states are held in different static trapping potentials, omitting the effects of splitting and recombination. These will be addressed in future work.

We assume that the $\pi/2$ pulses are resonant with the atomic transition, such that just after the second $\pi/2$ pulse the density matrix reads:
\begin{eqnarray} \label{rhotf}
\widehat{\rho} (t_f) & = & \sum_n p_n \left\{ p_n^a \left|n,a\right>\left<n,a\right| + p_n^b \left|n,b\right>\left<n,b\right| \right.\nonumber\\
  & + & \left. p_n^{ab}\left[ \left|n,a\right>\left<n,b\right| +\text{h.c.} \right]  \right\} \;,
\end{eqnarray}
where $p_n^a = \left[1-\cos{\left( \delta \omega_n t \right)}\right]/2$, $p_n^b = \left[1+\cos{\left(\delta \omega_nt \right)}\right]/2$ and $p_n^{ab} = \sin{\left(\delta \omega_nt\right)} /2$. Here, $\hbar \delta \omega_n = E_n^b-E_n^a$ is the difference between the corresponding energy levels in the two wells. The quantities $p_n^a$ and $p_n^b$ are the populations of levels $\left|n,a\right>$ and $\left|n,b\right>$, while the $p_n^{ab}$ are the coherence terms between $\left|n,a\right>$ and $\left|n,b\right>$. The total population in the internal state $\left|a\right>$ reads $p_a=\textrm{Tr} \left(\widehat{\rho} \left|a\right>\left<a\right|\right)$, leading, from equation~(\ref{rhotf}), to:
\begin{equation} \label{eqdepa}
p_a=\left(1/2\right)\left\{ 1 - \mathfrak{Re}\left[A(t_f)\right] \right\}\;,
\end{equation}
with $A(t) = \sum_n p_n \exp{\left(j \delta\omega_n t\right)}$. In equation (\ref{eqdepa}), we identify the contrast as:
\begin{equation}
C(t) = \left|A(t)\right| = \left| \sum_n p_n \exp{\left(j \delta \omega_n t\right)} \right| \;.
\label{ContrastGene}
\end{equation}
As can be seen in equation~(\ref{ContrastGene}), the contrast is equal to unity if the eigenvalues of $\widehat{H} \left|a \right> \left< a \right|$ and $\widehat{H} \left|b \right> \left< b \right|$ are the same, which corresponds to the ideal case of a perfectly symmetric atom interferometer.

A simple formula for the coherence time can be derived from equation~(\ref{ContrastGene}) in the case where the two potentials correspond to harmonic (one-dimensional) traps with slightly different frequencies $\omega_a \neq \omega_b$. Under the additional hypothesis $\hbar\omega_{a,b} \ll kT$, equation (\ref{ContrastGene}) leads to the following value for $t_c$ (defined as the time for the contrast to be reduced to $1/2$):
\begin{equation}
t_c = \sqrt{3} \hbar\omega / \left( \delta\omega kT \right) \;,
%t_c = \frac{1}{\left| \delta \omega \right|}\frac{\hbar\omega}{kT}
\label{tcdw}
\end{equation}
with $\omega = \left(\omega_a+\omega_b\right)/2$ and $\delta\omega = |\omega_a-\omega_b|$. Equation (\ref{tcdw}) is the main result of this section. It shows that $t_c$ increases with the degree of symmetry (measured by $\delta \omega/\omega$) and decreases with temperature, as expected intuitively. As an example, we obtain $t_c \simeq$~15~ms for a temperature of 500~nK and a degree of symmetry on the order of $\omega/\delta\omega \simeq $~10$^{3}$. Furthermore, equation (\ref{ContrastGene}) can be used to quantitatively analyze other defects, for example anharmonic potentials.

The simple model presented in this section illustrates the importance of symmetry to maintain the coherence of the interferometer. As already discussed in the introduction, this can be seen as an atomic equivalent of white light interferometry in optics, where the path length between the two arms of the interferometer has to be made smaller than the coherence length. This is the main motivation for introducing the following protocol, which aims to achieve symmetrical state-dependent potentials using microwave dressing with two different frequencies on an atom chip.

\section{Proposal of a symmetric configuration}

\subsection{Basic principle of the protocol}

\begin{figure}[t]
\centering
\includegraphics[width=0.48\textwidth]{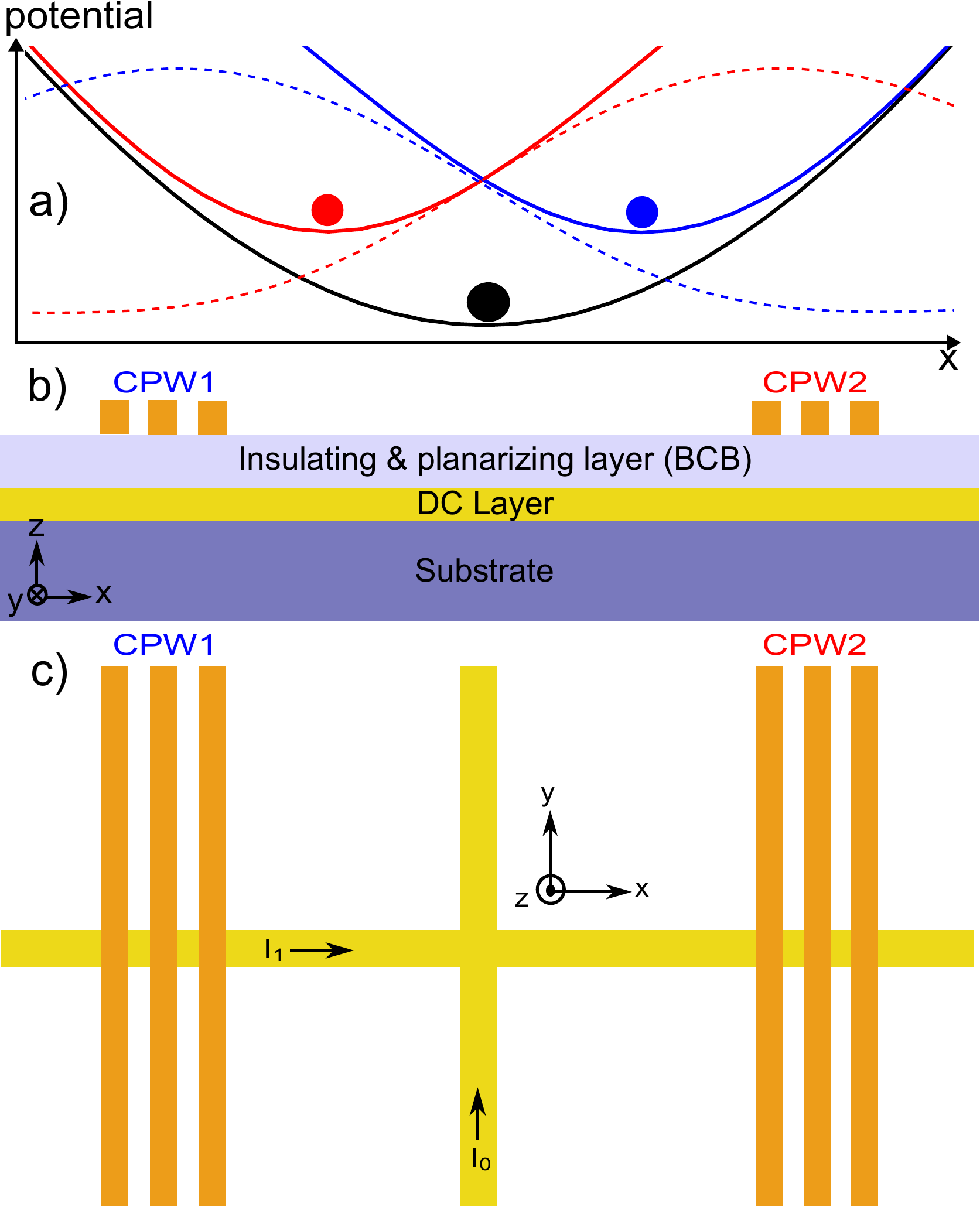}
\caption{Basic principle of state-selective symmetrical splitting. a)~Typical shapes of the adiabatic potentials in the near field of the coplanar waveguides (CPWs). b) Cut of the atom chip showing the CPWs and the DC layer, separated by an insulating and planarizing material. c) Top view of the atom chip. The central wires that carry the static currents $I_0$ and $I_1$ are used to create a static microtrap in the vicinity of the atom chip. The CPWs are deposited on both sides of the trap center, at equal distance from the central wire carrying $I_0$.}
\label{FigPuce}
\end{figure}

We consider in the following the experimental situation in which the $\ket{F=1, m_F=-1}$ and $\ket{F=2, m_F=1}$ hyperfine levels of the $5S_{1/2}$ ground state of $^{87}$Rb are used to implement the interferometric sequence described in the previous section (with $\ket{F=1, m_F=-1} \equiv \ket{a}$ and $\ket{F=2, m_F=1} \equiv \ket{b}$). These two states have nearly identical magnetic moments, rendering their superposition robust against magnetic field fluctuations \cite{harber2002effect}. The $\pi/2$ pulses described in the previous section are produced by two-photon (microwave and radio-frequency) pulses, as demonstrated in \cite{harber2002effect}. The potentials $V_a$ and $V_b$ are created by microwave dressing from two coplanar waveguides on the atom chip, as illustrated on figure~\ref{FigPuce}a. As already discussed in the introduction, this protocol is a generalization of \cite{bohi2009coherent} with two microwave frequencies on two separate coplanar waveguides (each one interacting mostly with one of the two states), with the goal of making them as symmetric as possible, as will be described in section \ref{SecSymBifreqSplitter}.

\subsection{Adiabatic dressed-state potentials}
\label{SecMWPotential}
%%%%%%%%%%%%%%%%%%%%%%%%%%%%%%%%%%%%%%%%%%%%%%%%%%%%%%%
%
In the presence of a DC magnetic field combined with a microwave field close to the hyperfine splitting frequency, the three Zeeman sublevels of $\ket{F=1}$ are coupled to the five Zeeman sublevels of $\ket{F=2}$, leading to ``dressed'' eigenstates \cite{dalibard1985dressed}.

Let us first consider the dressing on one two-level transition, where a state $\ket{F=1, m_1} \equiv \ket{g}$ is significantly coupled to only one state $\ket{F=2, m_2} \equiv \ket{e}$ by a microwave field with frequency $\omega$. The coupling strength is given by the Rabi frequency $\Omega$, which is proportional to the amplitude of the microwave magnetic field, and assumed to be much smaller than the Larmor frequency $\omega_L$ (i.e. the splitting with neighbouring Zeeman sublevels, given by $\omega_L=\mu_B B/(2\hbar)$, $B$ being the modulus of the DC magnetic field and $\mu_B$ the Bohr magneton) to ensure the validity of the two-level approximation. The energies of the resulting dressed states $\ket{\pm}$ are \cite{cohen1992processus}:
\begin{equation}
E_{\pm} = \frac{E_g+E_e}{2} \pm \frac{\hbar}{2}
\sqrt{\Omega^2 + \Delta^2} + \textrm{Constant} \;,
\label{dressed_states}
\end{equation}
where $E_g$ (respectively $E_e$) is the energy of the uncoupled level $\ket{g}$ (respectively $\ket{e}$), $\Delta=\omega-(E_e-E_g)/\hbar$ is the detuning, and the constant term accounts for the energy of the microwave field \cite{treutlein2008coherent}.

In the absence of any coupling ($\Omega=0$) the state $\ket{g}$ corresponds to the dressed state $\ket{+}$ or $\ket{-}$ depending on the detuning. As long as the coupling is varied adiabatically the atoms will remain in a single dressed state. The adiabatic condition reads \cite{schumm2005thesis}:
\begin{equation}
\vert \dot{\Omega} \Delta \vert \ll \left(\Delta^2 + \Omega^2 \right)^{3/2} \;.
\label{EqAdCondition}
\end{equation}
In this case, they will see the following adiabatic potential:
\begin{equation}
V_g = \frac{E_g+E_e}{2} + S_\Delta \frac{\hbar}{2}
\sqrt{ \Omega^2 + \Delta^2} - \frac{\hbar \omega}{2} \;,
\label{adiabatic_potential_g}
\end{equation}
where $S_\Delta$ is the initial sign of $\Delta$ (which we assume to be constant over the spatial extent of the atomic cloud). Similarly, the adiabatic potential for atoms initially in the bare state $\ket{e}$ reads:
\begin{equation}
V_e = \frac{E_g+E_e}{2} - S_\Delta \frac{\hbar}{2}
\sqrt{ \Omega^2 + \Delta^2} + \frac{\hbar \omega}{2} \;.
\label{adiabatic_potential_e}
\end{equation}
In equations (\ref{adiabatic_potential_g}) and (\ref{adiabatic_potential_e}), the average energy of the microwave field (in the sense of the semiclassical limit) has been removed, keeping only a $-\hbar \omega /2$ (respectively $+\hbar \omega /2$) term such that $V_g$ (respectively $V_e$) coincides with $E_g$ (respectively $E_e$) when $\Omega$ is initially set to zero.

%%%%%%%%%%%%%%%%%%%%%%%%%%%%%%%%%%%%%%%%%%%%%%%%%%%%ù
\subsection{Symmetric microwave dressing}
\label{SecSymBifreqSplitter}
\begin{figure}[t]
\centering
\includegraphics[scale=0.53]{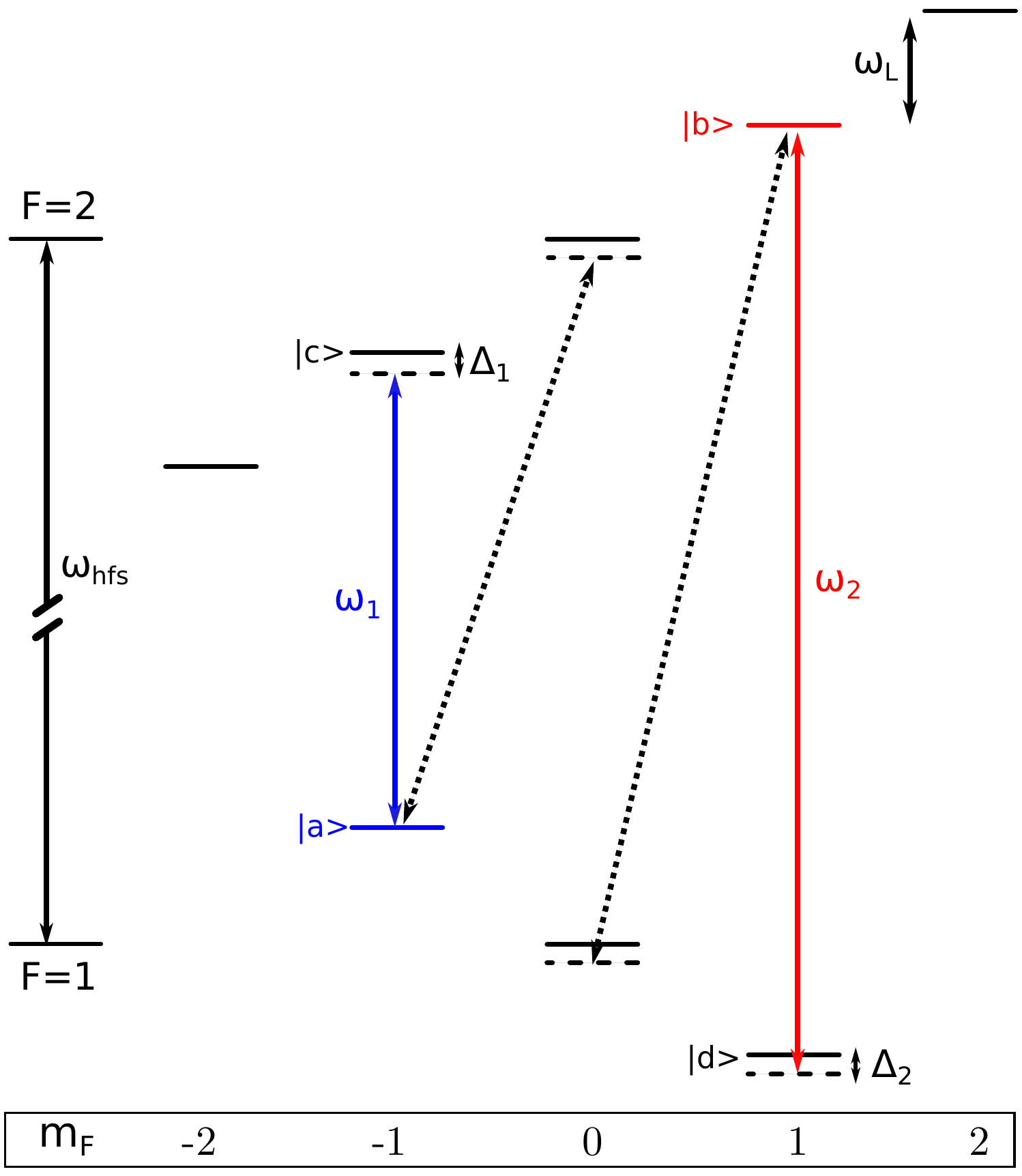}
\caption{Energy levels of the $5S_{1/2}$ ground state of $^{87}$Rb in the presence of a static magnetic field. To generate symmetrical state-dependent potentials, two microwave fields are used to couple the clock states $\ket{a}$ and $\ket{b}$ to two auxiliary states. Two combinations are possible by an appropriate choice of the microwave frequencies using either $\pi$ (solid line) or $\sigma$-transitions (dashed line). The $\pi$ (respectively $\sigma$) transitions correspond to the case where the microwave and DC magnetic fields are parallel (respectively orthogonal). Both configurations can be readily achieved for example using a regular dimple trap \cite{treutlein2008coherent}.}
\label{FigInternalStates}
\end{figure}

%%%%%%%%%%%%%%%%%%%%%%%%%%%%%%%%%%%%%%%%%%%%%%%%%%%%%
%
We now consider the situation in which two microwave frequencies are used to shift the energies of two pairs of levels, in order to achieve a microwave-induced, state-dependent potential. These two frequencies are injected into two different coplanar waveguides (labelled CPW$_1$ and CPW$_2$) placed on either side of the DC magnetic trap center, as sketched on figures~\ref{FigPuce}b and \ref{FigPuce}c. One possible implementation to make the potentials symmetric, illustrated on figure~\ref{FigInternalStates}, is to tune $\omega_1$ such that it is mostly resonant with the transition between $\ket{a}$ and $\ket{F=2, m_F=-1} \equiv \ket{c}$, while $\omega_2$ is tuned to be mostly resonant with the transition between $\ket{b}$ and $\ket{F=1, m_F=1} \equiv \ket{d}$. These conditions can be rewritten as $|\omega_1-(E_c-E_a)/\hbar|\ll \omega_L$ and $|\omega_2-(E_b-E_d)/\hbar|\ll \omega_L$, where $E_c$ (respectively $E_d$) is the energy of the bare state $\ket{c}$ (respectively $\ket{d}$), and $\omega_L$ is the Larmor frequency, defined in the previous section. If we furthermore assume that the amplitude of the microwave magnetic field is much smaller than $B$ (which means that all the Rabi frequencies corresponding to couplings between Zeeman sublevels of $F=1$ and $F=2$ are much smaller than $\omega_L$), then the two-level approximation can be used for the transitions $\ket{a} \leftrightarrow \ket{c}$ and $\ket{b} \leftrightarrow \ket{d}$. Following section \ref{SecMWPotential}, the adiabatic potential for the internal state initially in $\ket{a}$ then reads:
\begin{equation}
V_a = \frac{E_a+E_c}{2} - \frac{\hbar \omega_1}{2} + S_{\Delta_1} \frac{\hbar}{2} \sqrt{\Omega_1^2 + \Delta_1^2} \;,
\label{EqVa}
\end{equation}
where $\Delta_1=\omega_1-(E_c-E_a)/\hbar$ and $\Omega_1$ is the Rabi frequency associated with the transition $\ket{a} \leftrightarrow \ket{c}$ and the microwave field at frequency $\omega_1$. Similarly, the adiabatic potential for the internal state initially in $\ket{b}$ is:
\begin{equation}
V_b = \frac{E_b+E_d}{2} + \frac{\hbar \omega_2}{2} - S_{\Delta_2} \frac{\hbar}{2} \sqrt{\Omega_2^2 + \Delta_2^2} \;,
\label{EqVb}
\end{equation}
where $\Delta_2=\omega_2-(E_b-E_d)/\hbar$ and $\Omega_2$ is the Rabi frequency associated with the transition $\ket{b} \leftrightarrow \ket{d}$ and the microwave field at frequency $\omega_2$. The matrix elements of the interaction Hamiltonian associated to the transitions $\ket{a} \leftrightarrow \ket{c}$ and $\ket{b} \leftrightarrow \ket{d}$ are almost equal \cite{treutlein2008coherent}, which means that equivalent magnetic fields will lead to identical Rabi frequencies.

The energy of the bare states $\ket{a}$, $\ket{b}$, $\ket{c}$ and $\ket{d}$ can be approximated to the first order in $B$ (neglecting the coupling between the nuclear angular momentum and the magnetic field) by the usual Zeeman formula, namely $E_a= \hbar \omega_L$, $E_b= \hbar \omega_\textrm{hfs} + \hbar \omega_L$, $E_c= \hbar \omega_\textrm{hfs} - \hbar \omega_L$ and $E_d= -\hbar \omega_L$, where $\omega_\textrm{hfs}\simeq 2\pi \times 6.83$~GHz \cite{steck2001rubidium} is the zero-field hyperfine splitting (the common energy offset has been discarded). We furthermore impose that $\omega_1$ and $\omega_2$ be symmetrically tuned with respect to $\omega_\textrm{hfs}$, a condition which can be written as $\omega_1=\omega_\textrm{hfs} - \Delta_0$ and $\omega_2=\omega_\textrm{hfs} + \Delta_0$. This implies in particular that the initial detunings $\Delta_{1}$ and $\Delta_2$ have equal absolute values and opposite signs (we denote by $S$ the initial sign of $\Delta_1=2\omega_L-\Delta_0$). Equations (\ref{EqVa}) and (\ref{EqVb}) then read:
\begin{equation}
V_a (\textbf{r})= \frac{\hbar \Delta_0}{2} + S \frac{\hbar}{2} \sqrt{\Omega_1^2(\textbf{r}) + \left[2\omega_L(\textbf{r})-\Delta_0 \right]^2} \;,
\label{eqvabis}
\end{equation}
and $V_b(\textbf{r})=\hbar \omega_\textrm{hfs} + \tilde{V}_b(\textbf{r})$, with:
\begin{equation}
\tilde{V}_b (\textbf{r})= \frac{\hbar \Delta_0}{2} + S \frac{\hbar}{2} \sqrt{\Omega_2^2(\textbf{r}) + \left[2\omega_L(\textbf{r})-\Delta_0 \right]^2} \;.
\label{eqvbbis}
\end{equation}

Let us now consider the spatial dependence of $V_a$ and $V_b$ along the $x$-axis of figure \ref{FigPuce} in the framework of a simplified one-dimensional model. The DC magnetic trap is assumed to be harmonic and centered around $x=0$, such that $\omega_L(x)=\omega_L(-x)$. The two coplanar waveguides are assumed to be at the same distance on either side of the origin and fed with the same microwave power, such that $\Omega_1(x)=\Omega_2(-x)$ (recall that the interaction Hamiltonian has almost the same matrix elements for the two transitions). This leads to $V_a(x)=\tilde{V}_b(-x)$ which satisfies the desired symmetry condition. This is the main result of this section, showing that symmetry, in the sense defined in the previous section, is in principle possible with this configuration. This result can be generalized to the case of a more realistic geometry for the DC trap in three dimensions. In this case, the potentials $V_a$ and $V_b$ are found to be symmetric in the sense that they form two traps with similar eigenenergies.

One possible limitation of symmetry in this configuration is the presence of other (far off-resonance) transitions, although their effect is expected to be reduced at least by a factor on the order of $|\Delta_1|/\omega_L \ll 1$ as compared to the main $\ket{a} \leftrightarrow \ket{c}$ and $\ket{b} \leftrightarrow \ket{d}$ transitions, and can in principle be compensated by adjusting the power and frequency of the microwave dressing fields.

An alternative to the protocol described in this section is to use the $\sigma_+$ transitions $\ket{a}\leftrightarrow \ket{F=2, m_F=0}$ and $\ket{b}\leftrightarrow \ket{F=1, m_F=0}$ rather than $\ket{a} \leftrightarrow \ket{c}$ and $\ket{b} \leftrightarrow \ket{d}$, as illustrated by the dashed arrows of figure \ref{FigInternalStates}. We will not consider this alternative in detail in the following, but most of the results described in this paper can be transposed to it.
%
%%%%%%%%%%%%%%%%%%%%%%%%%%%%%%%%%%%%%%%%%%%%%%%%%

\section{Attractive versus repulsive microwave fields}
\label{SecBarrier}
%%%%%%%%%%%%%%%%%%%%%%%%%%%%%%%%%%%%%%%%%%%%%%%%%%
%
It can be seen in equations (\ref{eqvabis}) and (\ref{eqvbbis}) that when the initial sign $S$ of the detuning $\Delta_1$ is positive, both levels $\ket{a}$ and $\ket{b}$ will be blue-shifted: a maximum in the Rabi frequency $\Omega_{1,2}$ will result, for a constant value of the detuning $\Delta_1=2\omega_L-\Delta_0$, in a maximum of the adiabatic potential $V_{a,b}$ (as pictured in figure \ref{FigPuce}a). Consequently, the microwave field will be called ``repulsive'' in this case. In the opposite case ($S<0$), the microwave field will be called ``attractive''.

An important difference between repulsive and attractive microwave fields is the fact that the trap depth is limited in the latter case. This can be understood by first noticing that the Larmor frequency $\omega_L$ is minimal at the DC trap center, and increases with the distance from the center. In the attractive case, the detuning $\Delta_1=2\omega_L-\Delta_0$ is initially negative at the trap center, so it will go to zero for the points $\mathbf{r}$ in space corresponding to $\omega_L(\mathbf{r})=\Delta_0/2$, giving rise to an avoided crossing. Beyond this point, the magnetic dependence of the adiabatic potentials $\partial V_{a,b}/\partial B$ changes sign, and the atoms beyond this limit are no longer trapped by the DC field. This puts a limitation on the typical temperature that can be used in the attractive case, typically $kT \ll \hbar \Delta_0$. Conversely, in the repulsive case, the detuning $\Delta_1$ does not go to zero because it is initially positive at the trap center. The latter temperature constraint is thus relaxed.

A second reason to favor repulsive potentials arises from the fact that the atoms are trapped in a region of weaker microwave field than in the attractive case. This reduces the effect of the mixing of other atomic states in the trapped atoms. We discuss this effect in the next section.

%%%%%%%%%%%%%%%%%%%%%%%%%%%%%%%%%%%%%%%%%%%%%%%%%%%%%%%%%
\section{Robustness to magnetic field fluctuations}
\label{SecContamination}
%%%%%%%%%%%%%%%%%%%%%%%%%%%%%%%%%%%%%%%%%%%%%%%%%%%%%%%%%
%
In section \ref{SecSymBifreqSplitter}, we have approximated the hyperfine energy levels of $^{87}$Rb by the linear Zeeman formula, keeping only first order terms in $B \ll \hbar \omega_\textrm{hfs}/\mu_B$ and neglecting the coupling between the nuclear angular momentum and the magnetic field based on the fact that the electron spin g-factor is typically 3 orders of magnitude bigger than the nuclear spin g-factor. However, the latter is not negligible when superpositions of internal states are considered, because even a small difference in the magnetic dependence of the energy levels can strongly affect coherence in the presence of magnetic field noise. A remarkable situation occurs for the $\ket{F=1, m_F=-1}$ and $\ket{F=2, m_F=1}$ hyperfine levels of the $5S_{1/2}$ ground state of $^{87}$Rb (labelled $\ket{a}$ and $\ket{b}$ in this paper), whose energy difference is independent of $B$ to first order for a particular value $B_m \simeq 3.23$~G called the ``sweet spot'' \cite{harber2002effect,treutlein2004coherence}, making their coherent superpositions particularly robust to magnetic field fluctuations. In this section, we study the existence conditions for this sweet spot and, when applicable, the changes in the value of $B_m$ in the presence of microwave dressing.

To do this, we use the Breit-Rabi formula \cite{steck2001rubidium} for the hyperfine energy levels. Considering the fact that for most atomic physics experiments the magnetic field $B$ is typically much smaller than $\hbar \omega_\textrm{hfs}/\mu_B \simeq 0.5$~T, the energy levels for $F=1$ can be approximated up to the second order in $\mu_B B / (\hbar \omega_\textrm{hfs})$ by:
\begin{equation} \label{BR1}
E_{1,m_F}=\frac{m_F \mu_B B}{4}(5g_I-g_J)-\frac{\mu_B^2 \alpha (g_J-g_I)^2 B^2}{4 \hbar \omega_\textrm{hfs}} \;,
\end{equation}
where $\alpha=1-m_F^2/4$. Similarly, the energy levels for $F=2$ read $E_{2,m_F}=\tilde{E}_{2,m_F}+\hbar \omega_\textrm{hfs}$, with:
\begin{equation} \label{BR2}
\tilde{E}_{2,m_F}=\frac{m_F \mu_B B}{4}(3g_I+g_J)+\frac{\mu_B^2 \alpha (g_J-g_I)^2 B^2}{4 \hbar \omega_\textrm{hfs}} \;.
\end{equation}
In these formulas, $g_J \simeq 2.002$ and $g_I \simeq -9.95$~10$^{-4}$ are respectively the electron and the nuclear spin g-factors \cite{steck2001rubidium}. In the absence of microwave dressing, the usual sweet spot for $\ket{a}$ and $\ket{b}$ can be readily retrieved from equations (\ref{BR1}) and (\ref{BR2}) as the value of the magnetic field $B_m^0$ minimizing the energy difference $E_{2,1}-E_{1,-1}$, namely:
\begin{equation}
B_m^0 = \frac{-8 g_I \hbar \omega_\textrm{hfs}}{3 \mu_B (g_J-g_I)^2} \simeq 3.23 \, \textrm{G} \;.
\end{equation}
Let us now assume that we start from a situation with $B=B_m^0$ in the absence of microwave power, and that we then gradually ramp the Rabi frequencies $\Omega_1=\Omega_2$ up to a maximum value $\Omega$. The relevant energy levels (corresponding to the $\pi$ transitions of figure~\ref{FigInternalStates}) are then $E_a=E_{1,-1}$, $E_b=E_{2,1}$, $E_c=E_{2,-1}$ and $E_d=E_{1,1}$. It is convenient to specify the values of $\omega_{1}$ and $\omega_2$ via the initial detunings $\Delta_1^0=\omega_1-(E_c^0-E_a^0)/\hbar$ and $\Delta_2^0=\omega_2-(E_b^0-E_d^0)/\hbar$, where the notation $X^0$ refers to the value of $X$ at $B=B_m^0$. The problem can then be described by the two dimensionless parameters $\delta$ and $\kappa$, defined by:
\begin{equation}
\delta =  \Delta_1^0 / \omega_L^0 =-\Delta_2^0/ \omega_L^0 \quad \textrm{and} \quad \kappa = \left| \frac{\Omega}{\Delta_1^0} \right| \;,
\end{equation}
where $\omega_L^0 = \mu_B B_m^0/ (2 \hbar)$. Physically, $\kappa$ is linked to the degree of mixing in the dressed state picture \cite{dalibard1985dressed}. The initial sign $S$ of the detuning $\Delta_1$, as described in sections \ref{SecSymBifreqSplitter} and \ref{SecBarrier}, is equal in this case to the sign of $\delta$. The microwave field will be ``repulsive'' for $\delta >0$, and ``attractive'' in the opposite case. Equations (\ref{EqVa}) and (\ref{EqVb}) can be used to plot the energy difference $V_b - V_a$ as a function of $B$, for different values of $\delta$ and $\kappa$, and find the minimum when applicable.

\begin{figure}
\centering
\includegraphics[width=0.49\textwidth]{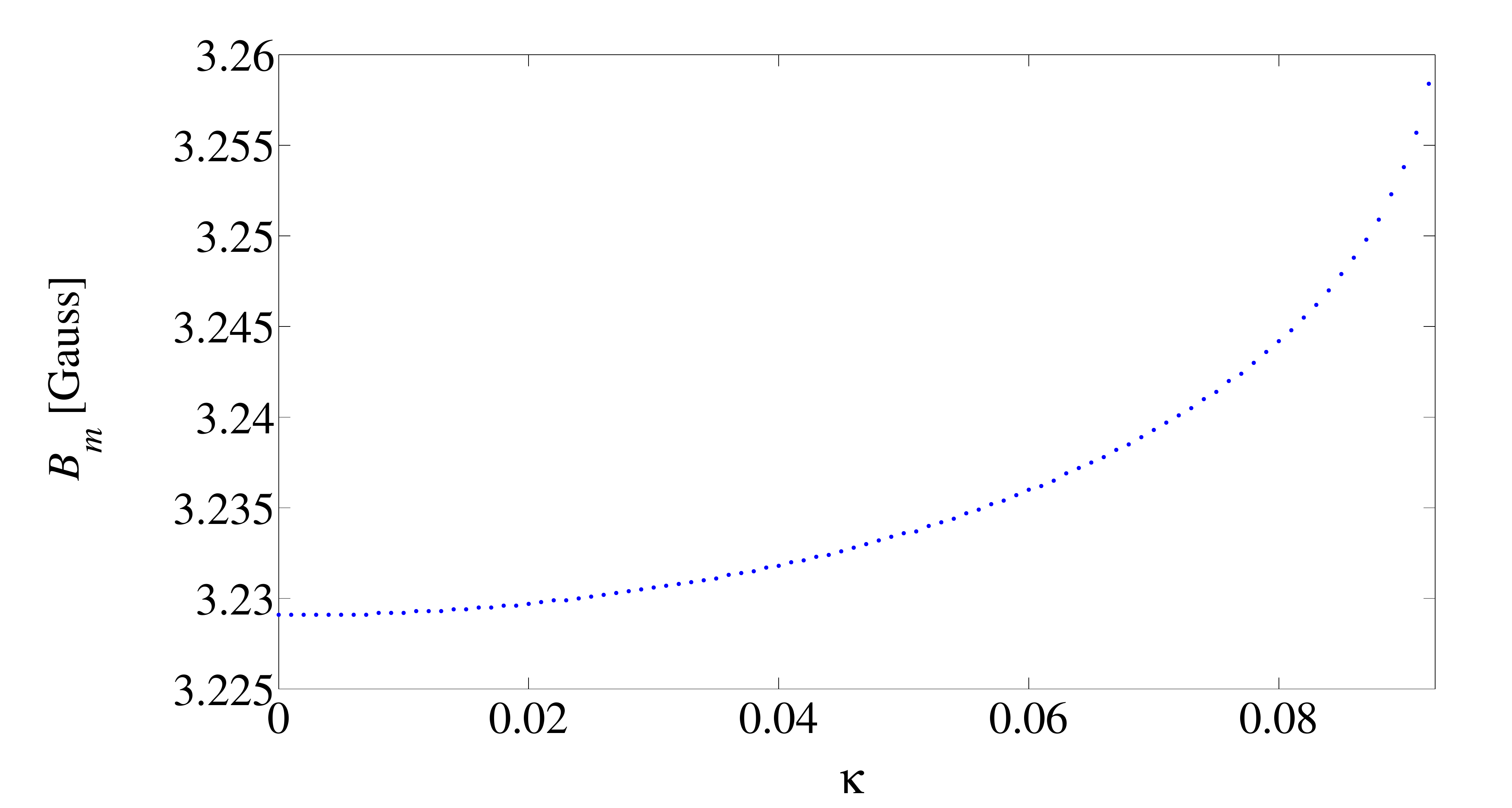}
\caption{Numerically computed value of the sweet spot $B_m$ (defined as the minimum of $V_b(B)-V_a(B)$) as a function of $\kappa$, with $\delta=-0.1$. The sweet spot remains up to a critical value on the order of $\kappa_c \simeq 0.092$.}
\label{delta_neg}
\end{figure}

\begin{figure}
\centering
\includegraphics[width=0.49\textwidth]{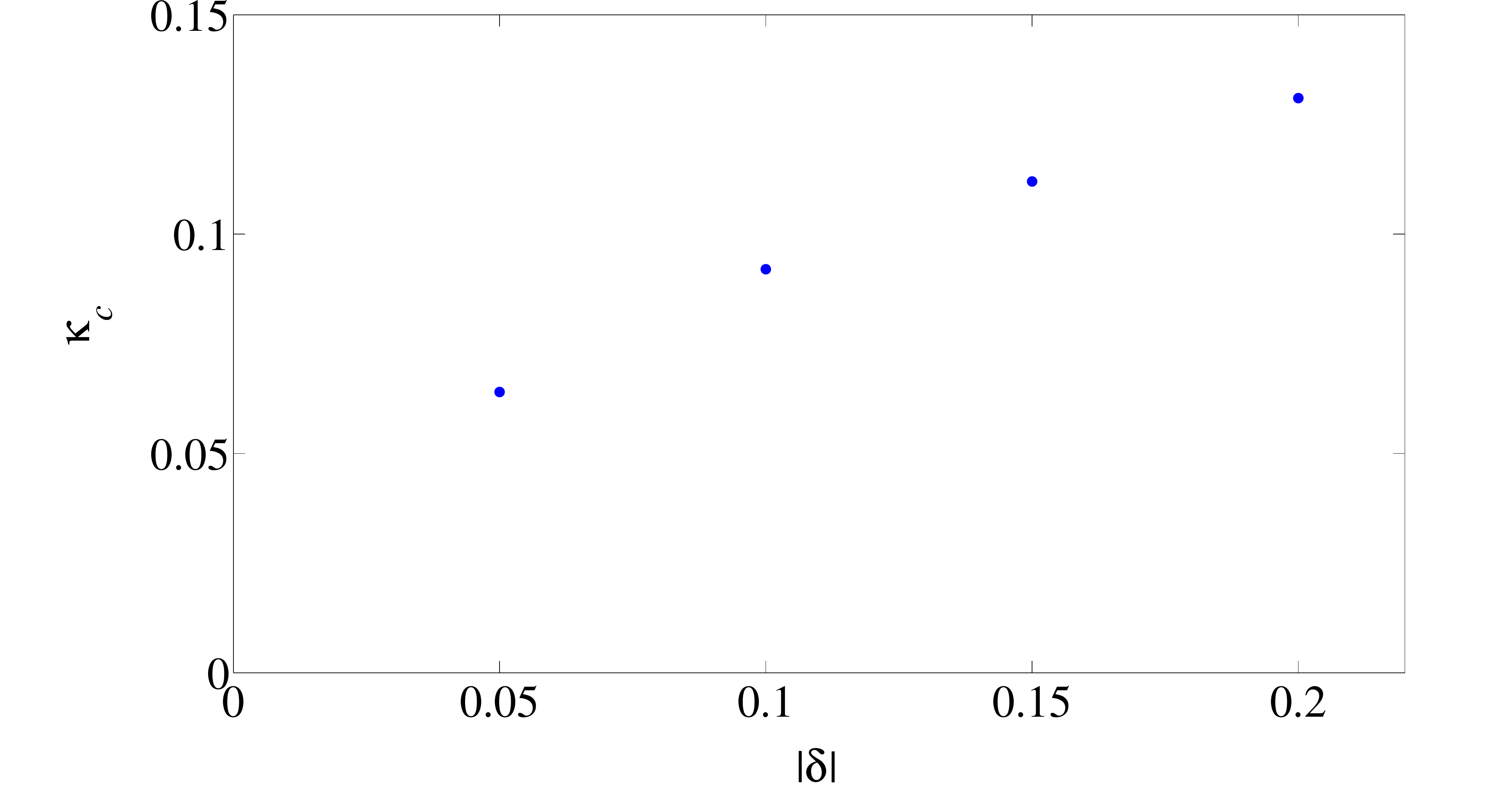}
\caption{Numerically computed value of critical value $\kappa_c$ (above which the sweet spot disappears) as a function of $|\delta|$, in the case $\delta<0$.}
\label{kappa_critic}
\end{figure}

In figure~\ref{delta_neg}, we show the case of an attractive microwave field by setting $\delta=-0.1$. In this case, we observe that the ``sweet spot'' value increases with $\kappa$, up to a critical value on the order of $\kappa_c \simeq 0.092$ where the minimum disappears. The value of $\kappa_c$ is observed to be a growing function of $|\delta|$, as illustrated on figure \ref{kappa_critic}. This will result, in the attractive case, in a trade-off between the maximum Rabi frequency that can be used and the minimum detuning of the microwave frequency.

\begin{figure}
\centering
\includegraphics[width=0.45\textwidth]{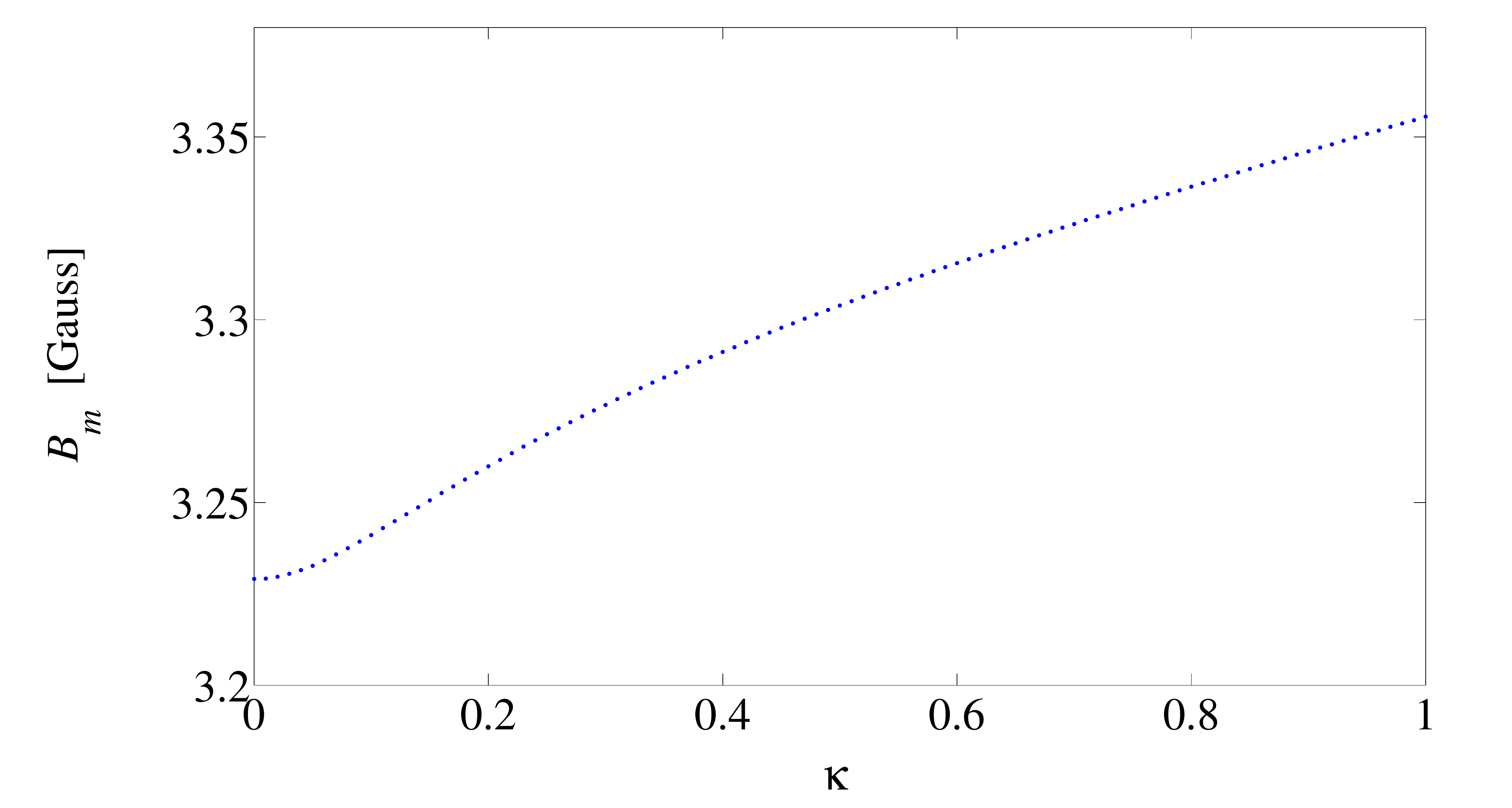}
\caption{Numerically computed value of the sweet spot $B_m$ as a function of $\kappa$, with $\delta=0.1$. }
\label{delta_pos}
\end{figure}

Let us now consider the opposite situation of a repulsive microwave field by setting $\delta=0.1$. In this case, a minimum of $V_b - V_a$ is found even for values of $\kappa$ much larger than unity, which is illustrated in figure \ref{delta_pos} for $0\leq \kappa \leq 1$. The situation remains the same for arbitrarily small values of $\delta>0$, which shows that the repulsive case is much more favorable than the attractive case, because it allows the Rabi frequency and the detuning to be chosen independently without compromising the existence of a sweet spot.

%%%%%%%%%%%%%%%%%%%%%%%%%%%%%%%%%%%%%%%%%%%%%%%%%%%%%%%ù

%%%%%%%%%%%%%%%%%%%%%%%%%%%%%%%%%%%%%%%%%%%%%%%%%%%%%%%%%%%%%%%%%%%%%ù

\section{Conclusion}

In conclusion, we have analyzed an experimental protocol for a symmetrical atom interferometer, based on the use of microwave dressing with two independent coplanar waveguides carrying different frequencies on an atom chip. We have pointed out the importance of symmetry for the contrast decay of a thermal atom interferometer in the framework of a simple model, and derived a simple formula for the coherence time in the harmonic case. This study shows that it is preferable to use a repulsive (rather than attractive) microwave field (i.e. $\delta>0$ with the notations used in this paper), because it avoids the problem of trap opening discussed in section \ref{SecBarrier}, reduces the degree of mixing $\kappa$ by confining the atoms in a region of weaker microwave fields, and ensures the existence of a ``sweet spot'' to reduce the sensitivity to magnetic field fluctuations even for strong microwave dressing fields.

A significant asset of this two-frequency protocol is that it provides independent control over the potentials seen by the two states. This feature gives additional degrees of freedom to counteract the residual dissymmetry, due for example to the effect of far off-resonance transitions that we have neglected in this paper.

Interferometry between internal states of thermal atoms on a chip has been shown to hold great promise for realizing compact cold atom clocks \cite{deutsch2010spin}. If experimentally successful, an atom chip interferometer with trapped thermal atoms could be an important step towards the achievement of a new class of compact integrated inertial sensors.

%%%%%%%%%%%%%%%%%%%%%%%%%%%%%%%%%%%%%%%%%%%%%%%%%%%ùù

\begin{acknowledgments}
The authors would like to thank A. Sinatra and I. Carusotto for useful discussions at the beginning of the project. This work has been carried out within the CATS project ANR-09-NANO-039 funded by the French National Research Agency (ANR) in the frame of its 2009 program in Nano-science, Nanotechnologies, and Nanosystems (P3N2009).
\end{acknowledgments}

\bibliographystyle{unsrt}

\end{document}